\documentclass{article}

\usepackage[bookmarks=true]{hyperref}

\hypersetup{
    colorlinks=true,
    linkcolor=black,
    filecolor=magenta,    
    citecolor=black,
    urlcolor=blue,
    pdfpagemode=FullScreen,
    }

\usepackage{graphicx,epsfig,color}
\usepackage{cite}
\usepackage[english]{babel}
\usepackage{amssymb,amsfonts,amsmath}
\usepackage{verbatim}
\usepackage{bbm,bbold,bm}
\usepackage{tikz}
\newcommand{\be}{\begin{equation}} \newcommand{\ee}{\end{equation}}
\newcommand{\bea}{\begin{eqnarray}} \newcommand{\eea}{\end{eqnarray}}
\newcommand{\beann}{\begin{eqnarray*}}  \newcommand{\eeann}{\end{eqnarray*}}
\newcommand{\bfig}{\begin{figure}} \newcommand{\efig}{\end{figure}}
\newcommand{\ba}{\begin{array}} \newcommand{\ea}{\end{array}}
\newcommand{\bcen}{\begin{center}} \newcommand{\ecen}{\end{center}}
\newcommand{\btab}{\begin{tabular}} \newcommand{\etab}{\end{tabular}}

\newtheorem{Proposition}{Proposition}[section]

\newtheorem{Theorem}{Theorem}[section]
\newtheorem{Lemma}{Lemma}[section]
\newtheorem{Corrolary}{Corrolary}[section]

\newcommand{\bp}{\begin{Proposition}}   \newcommand{\ep}{\end{Proposition}}
\newcommand{\bt}{\begin{Theorem}}   \newcommand{\et}{\end{Theorem}}
\newcommand{\bl}{\begin{Lemma}}     \newcommand{\el}{\end{Lemma}}
\newcommand{\bc}{\begin{Corrolary}} \newcommand{\ec}{\end{Corrolary}}
\newcommand{\chdens}{\rho}

\newcommand{\momdens}{p}

\newcommand{\eqendspace}{\,}

\title{\bf \Large Space-dependent symmetries and fractons\\ {\em \small Non-Lorentzian Geometry and its Applications}}

\author{
 Kevin T. Grosvenor${}^1$, Carlos Hoyos${}^2$, Francisco Pe\~{n}a-Ben\'{\i}tez${}^{3,4}$ and Piotr Sur\'owka${}^{3,4}$,\\
\small ${}^1$ {\em Instituut-Lorentz, Universiteit Leiden, P.O. Box 9506, 2300 RA Leiden, The Netherlands.} \\
\small ${}^2$ {\em Department  of  Physics  and  Instituto  de  Ciencias  y  Tecnolog\'{i}as  Espaciales  de  Asturias  (ICTEA)}\\
\small {\em Universidad  de  Oviedo,  c/  Federico  Garc\'{i}a  Lorca  18,  ES-33007  Oviedo,  Spain.}\\
\small ${}^3$ {\em Department of Theoretical Physics,}   \\
\small {\em Wroc\l{}aw  University  of  Science  and  Technology,  50-370  Wroc\l{}aw,  Poland}\\
\small ${}^4$ {\em  Max Planck Institute for the Physics of Complex Systems and}\\
\small {\em W\"{u}rzburg-Dresden Cluster of Excellence ct.qmat, 01187 Dresden, Germany.}\\
}

\date{}

\begin{document}

\maketitle

\begin{abstract}
There has been a surge of interest in effective non-Lorentzian theories of excitations with restricted mobility, known as fractons. Examples include defects in elastic materials, vortex lattices or spin liquids. In the effective theory novel coordinate-dependent symmetries emerge that shape the properties of fractons. In this review we will discuss these symmetries, cover the effective description of gapless fractons via elastic duality, and discuss their hydrodynamics.
\end{abstract}

\tableofcontents

\section{Introduction}

Fractons are usually identified with excitations of a system that are immobile or have restricted mobility - they can propagate along some spatial directions but not along others. In a pair of groundbreaking papers \cite{Pretko:2016kxt, Pretko:2016lgv},
Pretko demonstrated how the simultaneous conservation of charge and dipole moment (and even a particular component of the quadrupole moment) naturally leads to fractons, and he pioneered the study of symmetric tensor gauge theories as models containing gapless fractonic excitations. The gapless nature of the fractonic excitations in these symmetric tensor gauge theories contrasts against the gapped fractonic models that had been studied previously in lattice models (e.g., the Haah code~\cite{Haah2011} and the X-cube model~\cite{Vijay:2016phm}). 
Later on, Gromov initiated a systematic classification of fracton phases of matter based on symmetry principles privileging the charge and dipole symmetries and their higher-order generalizations -- the multipole algebra~\cite{Gromov:2018nbv}. He noted that the multipole algebra for a scalar field theory was on-the-nose the same as the so-called \emph{polynomial shift symmetries} that had been studied previously 
by Griffin, Grosvenor, Ho\v{r}ava and Yan (GGHY) in the context of technical naturalness in non-relativistic quantum field theories~\cite{Griffin:2013ub, Griffin:2015uz}. We will review how symmetric tensor theories emerge from the point of view of the realization of these symmetries, and discuss some physical systems where they appear.

In Sec.~\ref{sec:shift}, we shall discuss how the polynomial shift symmetries came about and how they are related to fractons. Our approach is based on symmetry principles and as such complements existing reviews on various aspects of fractons from a condensed matter perspective \cite{Nandkishore:2018sel,Pretko:2020cko}. This program has been established only recently and is far from complete. The best understood example consist of the symmetric tensor gauge theories, whose geometric nature can be understood in terms of the Heisenberg  symmetry group and the associated group manifold. We present this in Sec \ref{sec:Heisenberg}. This formalism should be viewed as a starting point for a more detailed analysis of low-energy theories containing fractons. In Sec.~\ref{sec:elastic} we discuss examples of such theories, which include the theory of elasticity and its generalizations. Following Pretko and Radzihovsky we show that symmetric elasticity in two spatial dimensions can be mapped via a duality transformation to the symmetric tensor gauge theories, whose geometric structure corresponds precisely to the Heisenberg group. We also show two examples of more general elastic theories that can be mapped to dual gauge theories with fractons: elasticity of quasi-crystals and Cosserat elasticity with internal, rotational degrees of freedom. A common feature of these theories is that fractons serve as charges. Finally, in Sec.~\ref{sec:hydro} we discuss some of the existing proposals to describe the collective dynamics of fractons in a low energy effective theory. 


\section{Fractons from polynomial shift symmetries} \label{sec:shift}

In this section, we will discuss the story behind the polynomial shift symmetries, their motivations and history, some of their consequences, and their link to fractons. These symmetries grew out of the work of GGHY on technical naturalness in non-relativistic quantum field theory. These authors conceived of these polynomial shift symmetries as allowing the existence of degrees of freedom (Nambu-Goldstone Bosons, in this case) with dispersion relations that were higher-order than linear or quadratic, which were the standards prior to that work~\cite{Watanabe_2012, Watanabe_2013}. While these authors were clearly fishing for condensed matter realizations of these degrees of freedom, their fractonic nature, that is, their restricted mobility, was not recognized at the time. 

The principle of technical naturalness is simply the conviction that the macroscopic behavior of systems follow from the equations governing their microscopic constituents. Even if it were impossible in practice to trace this relationship exactly from short distances, or high energies, to large distances, or low energies, it should nevertheless be possible as a matter of principle. The equations that govern this tracing from high to low energies are the celebrated renormalization group (RG) equations. Equations with parameters that are not finely tuned to be extremely small or large do not exhibit solutions that themselves contain finely tuned parameters. As articulated in 't~Hooft's 1979 paper~\cite{t1979naturalness}, only when there is a symmetry of the system whose exact preservation would forbid the existence of a certain parameter can that parameter be reasonably expected to take on such a small value. The classic example is the electron mass, which, at $m_e = 0.511$ MeV, is much smaller than the electroweak symmetry breaking scale, $v = 246$ GeV. This does not constitute fine tuning because chiral symmetry, namely the separate conservation of left- and right-chiral electrons, would forbid a nonzero electron mass entirely. In contrast, no such enhancement of symmetry occurs when the mass of a fundamental scalar field is set to 0, which is why we would not expect a fundamental scalar field to be very light and why it is puzzling that the Higgs mass should be so much smaller than the Planck scale. Naturalness questions are not restricted to particle physics, either. A famous example of a naturalness puzzle in condensed matter physics is the linear resistivity of strange metals: over a large range of temperatures above their critical temperature, the resistivity of high-$T_c$ superconductors scales linearly with the temperature \cite{Polchinski:1992ed}. Symmetry arguments for effective field theories that could describe the electrons and their interactions with themselves and their environment would lead one to expect scalings of the form $T^0$ (electron-impurity scattering), $T^2$ (electron-electron scattering), and $T^5$ (electron-phonon scattering). These symmetries are traced through the RG equations: a theory that enjoys a certain symmetry will not generate interactions under renormalization that break that symmetry. Put differently, the RG flows, or beta functions, of symmetry-breaking parameters vanish when these parameters themselves vanish.

Indeed, GGHY did not just cook up polynomial shift symmetries and then build theories based on them. The discovery went quite the other way round: they were studying the RG of the $z=2$ Lifshitz linear and non-linear $O(N)$ sigma models and noticed that the RG equations themselves protected the smallness of the $z=1$ kinetic term. For example, consider an $O(N)$ vector of scalar fields, $\boldsymbol{\phi}$, in $(3+1)$ dimensions, with quadratic terms given by 
\begin{align}
    \mathcal{L}_0 &= \frac{1}{2} \boldsymbol{\phi} \cdot \mathcal{O} \boldsymbol{\phi}, &%
    \mathcal{O} &= - \partial_0^{2} - \partial^4 + c^2 \partial^2 - m^4,
\end{align}
where $\partial_i$, $i=1, \ldots , d$  denotes a derivative with respect to spatial coordinates, with $\partial^2 \equiv \partial_i \partial_i$ and $\partial^4 = ( \partial^2 )^2$, and where we keep all relevant and marginal $O(N)$-invariant local interaction terms. The most relevant interaction is the quartic coupling
\begin{equation}
    \mathcal{L}_{\text{int}} = - \frac{\lambda}{4} \bigl( \boldsymbol{\phi} \cdot \boldsymbol{\phi} \bigr)^2.
\end{equation}
The leading one-loop correction to $c^2$ in the unbroken phase, in which $\boldsymbol{\phi} \cdot \boldsymbol{\phi}$ has zero vacuum expectation value (vev), comes from the two-loop diagram 
\begin{minipage}{1.4cm}
\begin{tikzpicture}[scale=0.4]
    \draw (-1.5,0) -- (-0.5,0);
    \draw (0.5,0) -- (1.5,0);
    \draw (-0.5,0) -- (0.5,0);
    \draw (0,0) circle [radius=0.5];
\end{tikzpicture}
\end{minipage}
and scales as $\delta c^2 \sim \lambda^2 / m^4$. Now, the constant shift symmetry $\boldsymbol{\phi} \rightarrow \boldsymbol{\phi} + \boldsymbol{a}$, where $\boldsymbol{a}$ is a constant vector, can force $\lambda$ and $m^4$ (the parameters that appear in the action) to be simultaneously small, say, $\lambda \sim \varepsilon \mu^3$ and $m^4 \sim \varepsilon \mu^4$, where $\varepsilon \ll 1$ and $\mu$ absorbs the dimensions of these parameters and sets the \emph{scale of naturalness}. Plugging these scalings into $\delta c^2$ gives
\begin{equation}
    \delta c^2 \sim \frac{\lambda^2}{m^4} \sim \frac{\varepsilon^2 \mu^6}{\varepsilon \mu^4} = \varepsilon \mu^2,
\end{equation}
which immediately raises the question of why the correction to the speed term should be small even though it is \emph{not} killed by the constant shift symmetry. The answer is clear: there must be another symmetry at work that kills all three terms $\lambda$, $m^4$, and $c^2$ simultaneously. This is the quadratic shift symmetry, in which the shift parameter $\boldsymbol{a}$ is promoted from a constant to a general quadratic polynomial of the spatial coordinates, $\boldsymbol{a} = \boldsymbol{a}_{ij} x^i x^j + \boldsymbol{a}_i x^i + \boldsymbol{a}_0$.  Thus, the polynomial shift symmetries were discovered.

This rather innocuous looking observation has some deep and surprising consequences. For example, the prevailing wisdom at the time was that Nambu-Goldstone bosons (NGBs) arising from spontaneous symmetry breaking could only stably exist with linear and quadratic dispersion relations~\cite{Watanabe_2012, Watanabe_2013}. In contrast, the polynomial shift symmetries allowed for the existence of NGBs with higher-order dispersion relations. Furthermore, complicated patterns of symmetry breaking lead to hierarchies of different powers of dispersion along the RG flow from high to low energies~\cite{Griffin:2015tu}. This latter observation leads to an extension of the celebrated Coleman-Hohenberg-Mermin-Wagner (CHMW) theorem, which forbids the spontaneous breaking of continuous internal symmetries in equilibrium field theories and their corresponding NGBs in dimensions one and two \cite{hohenberg, Mermin_Wagner, Coleman:1973ci} (see also Halperin's discussion of the history of this theorem as well as its limitations \cite{Halperin_2018}). In condensed matter, this famously kills long-ranged order (e.g., of ferromagnetic or anti-ferromagnetic kind) in lattice systems of dimension one or two. This extends to NGBs whose kinetic term is quadratic in time derivatives (so-called Type-A NGBs), which cannot exist as stable objects in spatial dimensions $d \leq z$, where $z$ is the dynamical critical exponent appearing in the dispersion relation $\omega^2 \propto ( \boldsymbol{k} \cdot \boldsymbol{k} )^z$. On the other hand, Type-B NGBs, whose kinetic term is linear in time derivatives, are immune from the CHMW theorem and can exist in any dimension~\cite{Griffin:2015tu}. 

Another intriguing consequence of these symmetries was found when Gromov made the link between polynomial shift symmetries and the restricted mobility of fractons~\cite{Gromov:2018nbv}. A simple example of this is a scalar field theory with a Lagrangian that is a functional of $\partial_0 \phi$ and $\partial_i \partial_j \phi$ (see \cite{Grosvenor:2021rrt} for the more general case when the Lagrangian depends on $\partial_i \phi$ as well as higher spatial derivatives of $\phi$). Such a theory enjoys the linear shift symmetry in the spatial coordinate $\phi \rightarrow \phi + \alpha + \beta_i x^i$, where $\alpha$ and $\beta_i$ are constants. The Noether currents of these symmetries are given by
\begin{subequations}
\begin{align}
    J_{\alpha}^{0} & \equiv \rho = \frac{\partial \mathcal{L}}{\partial ( \partial_0 \phi )}, &%
    J_{\alpha}^{i} &= - \partial_j \frac{\partial \mathcal{L}}{\partial ( \partial_i \partial_j \phi )} \equiv \partial_j J^{ij}, \\
    J_{\beta}^{i0} &= \rho x^i, &%
    J_{\beta}^{ij} &= x^i J_{\alpha}^{j} - J^{ij}.
\end{align}
\end{subequations}
The conservation equation $\partial_{\mu} J_{\beta}^{i \mu} = 0$ follows directly from the conservation equation $\partial_{\mu} J_{\alpha}^{\mu} = 0$, which itself is simply the Euler-Lagrange equation of motion and reads 
\begin{equation}
    \partial_0 \rho + \partial_i\partial_j J^{ij} = 0.
\end{equation}
A direct consequence of this is, of course, the conservation of the total charge and dipole moment:
\begin{align}
    Q &\equiv \int_{\Sigma} \rho, &%
    Q^i &\equiv \int_{\Sigma} \rho x^i,
\end{align}
where $\Sigma$ is any spatial slice in spacetime. That is, $\frac{dQ}{dt} = 0$ and $\frac{d Q^i}{dt} = 0$. These conservation laws prohibit the motion of solitary particles, but allows for the motion of composite particles, for example a solitary dipole made of one positive and one negative charge. Higher-order polynomial shifts lead to conservation laws for higher order multipoles and more intricate restrictions on the mobility of the degrees of freedom. This is how polynomial shift symmetries lead to fractonic excitations and this realization has promoted these symmetries to one of the guiding principles behind many of the recent studies in fractons.


\section{Polynomial shift symmetry and the Heisenberg Algebra} \label{sec:Heisenberg}

As was discussed in the previous section, polynomial shift symmetries can be associated with the conservation of a set of multipole charges in the corresponding system. In fact, it has been argued that this class of theories show fractonic behavior. More precisely, they fit within the gapless phases \cite{Xu2006,Xu2010,Pretko:2016kxt, Pretko:2016lgv,You_emergent_2020,Pretko_Zhai_2019}. 

The simplest case corresponds with the conservation of a charge $Q$ and its dipole $Q^i$, which in $d$ space dimensions, at the macroscopic level, can be formulated in terms of  a charge density $\rho$ as  
\begin{align}
	\frac{d Q}{dt} &=\frac{ d}{ dt}\int d^d x\,\rho =0 \,,\label{eq_monopole}\\
	\frac{ d Q^i}{ dt} &=\frac{ d}{ dt}\int d^d x\,  x^i\rho\,=0\,.\label{eq_dipole}
\end{align}
In a system with such conservation law, charges are immobile, whereas dipoles can freely move.
In fact, similarly to what happens with momentum and angular momentum
, both charges are conserved once the single (generalized) continuity equation 
\begin{equation}\label{eq_scalconserv}
	\partial_0\rho+\partial_i\partial_j J^{ij}=0\,,\qquad i,j=1,2,\ldots,d\,,
\end{equation}
is satisfied. The distinguishing feature in these class of systems is that charge is relaxed via a tensorial current. An immediate consequence of such conservation law, is that a gauged version of the symmetry would require the presence of gauge fields $A_0,A_{ij}$ with the transformation rule $A_0\to A_0-\partial_0\alpha$, and $A_{ij}\to A_{ij}+\partial_i\partial_j\alpha$, and the 'gauge fields' coupling to the fractonic matter as follows
\begin{equation}
	S = S_0[A_0,A_{ij}]+\int\,d^{d+1}x\left(\rho A_0 + J^{ij}A_{ij} \right)\,.
\end{equation}
Such type of theories have been proposed as a generalization to electrodynamics \cite{Pretko:2016lgv,ShenoyMoessner2020}. However, due to the unusual transformation law of the fields, it is not clear in what sense they are gauge theories. In addition, from this perspective, is not obvious whether it is possible to put the theory on a curved manifold without spoiling the gauge symmetry \cite{slagle2019symmetric}. In the recent paper \cite{Pena-Benitez:2021ipo}, it was pointed out that fracton gauge transformations are actually spacetime transformations and in order to properly implement them, they must be treated as such. To illustrate this fact, let us suppose a translational invariant system with monopole and dipole conservation. In a classical low energy regime the Poisson brackets between the charge $\rho$ and momentum $p_a$ densities are
\begin{subequations}
	\begin{align}
		[ \momdens _ i(\bm{x}) , \chdens (\bm{y})    ] &= - \chdens (\bm{x}) \partial _{x^i} \delta(\bm{x}-\bm{y}) \eqendspace,\\
		[ \momdens _ i(\bm{x}) , \momdens_j(\bm{y})    ] &= -[ \momdens_j(\bm{x}) \partial _{x^i} 
		+ \momdens_i (\bm{y}) \partial _{x^j} ] \delta(\bm{x}-\bm{y}) \eqendspace,
	\end{align}\label{eq:Poisson}
\end{subequations}
which imply the following non-vanishing bracket between the dipole charge and momentum
\begin{align}\label{eq_Halgebra}
	[P_i,Q^j] &= \delta_i^j Q\,,
\end{align}
where $P_i$ is the generator of spatial translations. The non-vanishing of  Eq. \eqref{eq_Halgebra} implies the fractonic transformations and space translations form a non-Abelian group. This fact will have important consequences in how the symmetry is realized, and the amount of  Nambu-Goldstone bosons  when the fracton symmetry is spontaneously broken. Actually, after carefully analyzing Eq. \eqref{eq_Halgebra} we notice the similarity with the canonical commutation relation in quantum mechanics between position and momentum. This algebra has been extensively studied in mathematics, and operators satisfying such commutation relation generates a Lie group called Heisenberg group \cite{calin2005geometric,vukmirovic2015classification}. Below we will discuss some important properties of such a group and its relevance in the context of monople-dipole-momentum preserving theories.

\subsection{Heisenberg group}

The Heisenberg group is a Lie group generated by exponentiation  of  elements of the Lie algebra Eq. \eqref{eq_Halgebra}. For example, given an element $X=\boldsymbol x\cdot\boldsymbol  P + \boldsymbol z\cdot \boldsymbol Q + \phi Q$ of the algebra, a generic element of the group can be written as
\begin{equation}
	\mathcal G_X=e^{\boldsymbol x\cdot\boldsymbol  P}e^{\boldsymbol z\cdot\boldsymbol  Q}e^{\phi Q}\,.
\end{equation}
With this parametrization the left action of the group $\mathcal G_\xi\cdot\mathcal G_X = \mathcal G_{X'}$, with $\xi= \boldsymbol \zeta\cdot\boldsymbol  P + \boldsymbol \beta\cdot \boldsymbol Q + \alpha Q$, produces a new element with coordinates
\begin{equation}
	X' = (\boldsymbol x+\boldsymbol\zeta)\cdot\boldsymbol  P + (\boldsymbol z+\boldsymbol\beta)\cdot \boldsymbol Q + (\phi + \alpha -\boldsymbol{\beta}\cdot\boldsymbol x) Q\,.
\end{equation}
The Maurer-Cartan form  can be written as $\Theta=\Omega^{-1}d\Omega= v Q + e^iP_i + \omega_i Q^i$ with 
\begin{equation}
v = d\phi  + \boldsymbol z\cdot d\boldsymbol  x\,,\qquad 	e^i= dx^i\,,\qquad \omega_i = dz_i\,.
\end{equation}
In fact, notice that the one-forms $e^i,\omega_i,v$ are invariant under the left action of the group.  The Maurer-Cartan equations imply
\begin{equation}\label{eq_MCeq}
  dv = \omega_i\wedge e^i\,,\qquad	de^i=0\,,\qquad d\omega_i =0\,\,.
\end{equation}

This parametrization of the group is usually known in the mathematics literature as polarized Heisenberg group. This group  has a unitary representation which acts projectively in the Hilbert space, 
given a non-zero real number $q$ the group acts as
\begin{equation}
	\pi_q[\mathcal G_\xi]\psi(t,\boldsymbol x) = e^{iq\alpha  }e^{i\boldsymbol\beta \cdot \boldsymbol x}\psi(t,\boldsymbol x +\boldsymbol \zeta)\,,
\end{equation}
where $q$ can be interpreted as the elementary fracton charge, $\alpha ,\boldsymbol \beta$ parametrizing the generalized $U(1)$ fracton global transformation, and $\boldsymbol{\zeta}$ a space translation respectively.

Since the group is a Lie group, it can be identified with a differential manifold that we denote as $\mathcal N_{2d+1}$ with coordinates $(x^i,z_i,\phi)$. In addition, the manifold posses  the non-commutative operation
\begin{equation}
	(x^i,z_i,\phi) \to (x^i +\zeta^i ,z_i + \beta_i , \phi + \alpha -\beta_i x^i)\,,
\end{equation}
as isometry group. Given this property, we can define the invariant basis
\begin{align}
	\mathcal	P_i  &= \frac{\partial}{\partial x^i} - z_i\frac{\partial}{\partial \phi}\,,\qquad e^i =dx^i\,,	\\
	\mathcal	Q^i  &= \frac{\partial}{\partial z_i} \,,\qquad\qquad\qquad \omega_i = dz_i\,, \\
	\mathcal	Q &= \frac{\partial}{\partial \phi}\,,\qquad\qquad\qquad v = d\phi + \boldsymbol z\cdot d\boldsymbol  x\,.
\end{align}
These vector fields satisfy the Lie bracket
\begin{equation}
	[\mathcal P_i,\mathcal Q^j] = \mathcal J(\mathcal P_i,\mathcal Q^j)\mathcal Q =\delta^j_i \mathcal Q \,,
\end{equation}
which agrees with the Lie algebra Eq. \eqref{eq_Halgebra}. Notice that in the previous formula we have introduced the `symplectic' form $\mathcal J=-dv$ which can be written as
\begin{equation}\label{eq_simplform}
	\mathcal{J} = e^i\wedge \omega_i\,.
\end{equation}
Therefore, the symplectic form encodes the non-trivial part of the Maurer-Cartan equation (see Eq. \eqref{eq_MCeq}) which accounts for the infinitesimal properties of the Heisenberg group. From this perspective, it is not surprising that the Lie bracket between the basis vector fields is given by $\mathcal J$.

\subsection{From the Heisenberg space to the fractonic system}

In contrast to the usual case of internal symmetries, the actual physical space is non-trivially connected with the `internal' space of fracton transformations. Therefore we will assume that the physical spacetime  is embedded into the Heisenberg space. To do so, we first extend $\mathcal N_{2d+1}\to \mathcal N_{2d+1}\times \mathbb{R}$, assuming the extra coordinate is the time, which we refer   to here as $x^0$, and include time translations into the isometry group of spacetime. 

After such extension, we have the extra basis vector and co-vector $\mathcal H=\partial/\partial x^0$, and $\tau=dx^0$ respectively, and can define the invariant metric\footnote{In this section we will assume rotational invariance.}
\begin{equation}
	\bar	G = s\tau^2 + e^ie^i +\omega_i\omega_i  + v^2 \,,
\end{equation}
with $s$ a sign that will be fixed below. 

Notice that the extended Heisenberg space must be understood as an abstract space containing both the physical spacetime with coordinates $(x^0,x^i)$, and the internal `generalized $U(1)$' directions $(z_i,\phi)$. Therefore, when the symmetry is not spontaneously broken, the points in the `internal' directions should be identified as $z_i\sim z_i +\beta_i$, and $\phi\sim \phi+\alpha -\beta_i x^i$, implying that the physical spacetime metric is
\begin{equation}
	G =	 s \tau^2 + e^i e^i  \,.
\end{equation}

On the other hand, when the internal symmetry is spontaneously broken, the internal directions become fields depending on the spacetime coordinates, and receive the interpretation of the Nambu-Goldstone bosons \cite{Pena-Benitez:2021ipo}. In fact, within our geometric interpretation, such a phase could be seen as having the physical spacetime embedded into the larger Heisenberg space with coordinates $(x^\mu,z_a(x^\mu),\phi(x^\mu))$, where the greek index $\mu$ refers to spacetime coordinates and takes values $0,1,\dots,d$. However, using Eq. \eqref{eq_simplform} and  the fact that $\mathcal J[\mathcal	P_i,\mathcal	P_j]=0$, we then conclude that $v$ on the physical spacetime has to have the form $v =\partial_0\phi \, \tau$, and the embedding of $z_i$ must satisfy the constraint $z_i=-\partial_i\phi$. After imposing these constraints we obtain
\begin{equation}
	\omega_i=-\partial_0\partial_i\phi\tau - \partial_i\partial_j\phi e^j\, .
\end{equation}
Actually, this reduction in the number of Nambu-Goldstone bosons is generically common when spacetime symmetries are spontaneously broken, and it is known as the inverse Higgs constraint \cite{isham1971nonlinear,low2002spontaneously,kharuk2018solving}.

In this case, the induced metric on the physical space differs from the metric $G_{\mu\nu}$ of a flat space by an extra symmetric tensor $B_{\mu\nu}$ depending on derivatives of the Nambu-Goldstone boson
\begin{equation}
	B = (\partial_0\phi \, dx^0)^2 + (\partial_0\partial_i\phi \, dx^0 +\partial_i\partial_j\phi \, dx^j)^2\,,
\end{equation}
and we refer to it as the fracton metric. This allows us to interpret the Goldstone mode as the breathing mode of the spacetime in the larger Heisenberg space.

Since we have constructed the proper invariants of the system,  a generic low-energy effective action for the spontaneously broken phase must have the form
\begin{equation}\label{eq_SSB1}
	S_{SSB}=\int d^{d+1}x\,\mathcal L\left(v_0,(\omega_{0i})^2,(\omega_{ij})^2\right)\,.
\end{equation}
In general, the form of  the effective Lagrangian would depend on the precise microscopic system we consider. However,  we notice that the Born-Infield action
\begin{equation}\label{eq_SSB2}
	S_{SSB}=-\int d^{d+1}x \sqrt{|G+B|}\,,
\end{equation}
corresponding with the volume of spacetime in the Heisenberg space  will guarantee a geodesic embedding. Using the derivative expansion $\partial_0 \sim \nabla^2$, and  expanding Eq. \eqref{eq_SSB2} up to a quadratic order  we obtain
\begin{equation}
	S_{SSB} \approx -\int d^{d+1}x\,\sqrt{|G|}\left(1 + \frac{s}{2}(\partial_0\phi)^2 + \frac{1}{2}(\partial_i\partial_j\phi)^2 + \ldots \right)  \,,
\end{equation}
notice that to guarantee a positive definite energy in this theory we must fix $s=-1$. 

This construction allows us to give a geometric interpretation to the Nambu-Goldstone boson associated to the spontaneous symmetry breaking in systems with monopole and dipole charge conservation, and naturally predicts the linear shift transformations of the gapless low energy mode $\phi$. Notice also that in \cite{Geng:2021cmq}, Geng, Kachru, Karch, Nally and Rayhaun studied fracton theories from brane constructions. Nonetheless, there is no obvious relation between their models and the Heisenberg space picture discussed here.

So far we have discussed the case of a global symmetry. However, certain spin liquids \cite{Pretko:2016lgv,You_emergent_2020,Pretko_Zhai_2019}, and elastic fields \cite{Pretko:2017kvd,Gromov:2018nbv,Nguyen:2020yve,du2021volume} contain a gauged version of the symmetry discussed here. In fact, a generalization of electrodynamics \cite{Pretko:2016lgv}  to account for systems with the class of  conservation laws discussed here has been constructed. In particular the monopole-dipole-momentum conserving case is described with gauge fields $A_0,A_{ij}$ with gauge transformations
\begin{equation}\label{eq_gaugetrans}
	\delta A_0= - \partial_0\alpha\,, \qquad \delta A_{ij} = \partial_i\partial_j\alpha\,.
\end{equation}
Nonetheless, if we were to allow the gauge fields to propagate in curved space, the gauge principle seems to enter in conflict with diffeomorphism transformations \cite{pretko2017emergent,Gromov:2018nbv}. In \cite{slagle2019symmetric} Slagle, Prem, and Pretko argued that in two space dimensions the fracton gauge invariance will not be broken as long as the space is an Einstein manifold, and more recently \cite{Nguyen:2020yve,du2021volume} the fractonic gauge symmetry has been extended to volume preserving diffeomorphisms and connected to the lowest Landau level.

In fact, by considering the structure of Heisenberg space Pe\~{n}a-Ben\'itez in \cite{Pena-Benitez:2021ipo} used standard techniques to gauge spacetime symmetries \cite{ivanov1982gauge,grignani1992gravity,son2006general,andringa2011newtonian} and managed to obtain a fully diffeomorphism- and gauge-invariant action for monopole-dipole-momentum conserving systems. One interesting prediction of  these analysis is that generically the monopole-dipole-momentum symmetry group  will be spontaneously broken once the system lives on an arbitrary curved spacetime.

Given a torsionless spacetime with frame fields  $\tau,e^i$ satisfying
\begin{equation}
	d\tau = 0\,,\qquad de^i -\omega^{ij}\wedge e^j =0\,,
\end{equation}
where $\omega^{ij}$ is the spin connection, and the metric is $G= -\tau^2 +e^ie^i$ the fracton field strength must be defined as
\begin{equation}
	F_i = D\tilde\omega_i\,,
\end{equation}
where $D$ is the rotational covariant exterior derivative, defined as $Ds^i=ds^i-\omega^{ij}\wedge s^j$. The $1-$form $\tilde\omega_i$ is related to the scalar and symmetric gauge fields via the relation
\begin{equation}
	\tilde\omega_i = -i_{\mathcal P_i}dA_0\tau + A_{ij}e^j\,.
\end{equation}
Fracton gauge transformations act on this field as $\delta\tilde\omega_i  = Di_{\mathcal P_i} d\alpha  $. However,  if the Riemann curvature tensor does not vanish, the non-Abelian structure of the symmetry group implies $\delta F_i\neq 0$. Therefore, in order to get a fully invariant theory it is necessary to introduce a Stueckelberg field $\phi$ transforming as $\delta\phi=\alpha$, and construct the invariant combination
\begin{equation}
	T_i = F_i -i_{\mathcal P_j}d\phi R^j\,_i\,,
\end{equation}
with $R^j\,_i$ the curvature $2-$form. After allowing for the spontaneous breaking of the symmetry, and defining the volume form $^*1=d^{d+1}x\,\sqrt{|G|}$, the following action can be constructed
\begin{equation}\label{eq_gaugeAction}
	S = -\frac{1}{2}\int\,{}^\star T_i\wedge T_i +  S_{SSB}[Di_{\mathcal P_i}d\phi-\tilde\omega_i,i_{\mathcal H}d\phi+A_0]\,.
\end{equation}

This theory satisfactorily reduces to the generalized electrodynamics theory of \cite{Pretko:2016lgv} in the flat space limit
\begin{align}
	S &=  \int d^{d+1}x\left[ F_{0 ij}F_{0ij} - \frac{1}{2}F_{ijk}F_{ijk}\right]\,.
\end{align}
In particular, in two space dimensions we can dualize the magnetic field and define electric and magnetic fields $E_{ij} =  F_{0 ij}$, and $B_k=\frac{1}{2}\epsilon^{ij}F_{ijk}$ respectively. The electromagnetic fields written in terms of the gauge potentials are
\begin{align}
	E_{ij} &= \partial_0 A_{ij} +\partial_i\partial_jA_0\,,\qquad	B_{k}  = \epsilon^{ij}\partial_i A_{jk} \,,
\end{align}
and the flat space gauge transformations are given by Eqs. \eqref{eq_gaugetrans}, as expected. In the next section we will study the relevance of this class of 'gauge theories' in the context of elasticity via the so-called Fracton-Elasticity duality.


\section{Fractons in elastic duals} \label{sec:elastic}

We have seen how tensor gauge fields appear naturally in the context of theories with polynomial shift symmetries. In the following we will discuss the physical relevance of these theories, with a view on condensed matter systems where these kind of ideas have had a larger impact. Experimental platforms of gapped type I fractons, with restricted mobility, has been suggested in \cite{Pretko_curcuit,Sous2020,Giergiel2021} and a proposal to realize immobile type II fractons has been put forward in \cite{Myerson-Jain}. A step towards a physical system with gapless fractons has been achieved when Pretko and Radzihovsky realized that a tensor gauge theory in two dimensions can be mapped to a familiar theory of elasticity where the fractonic excitations correspond to topological defects \cite{Pretko:2017kvd}. This discovery has been based on the earlier works of Kleinert that pioneered the field of elastic dualities in the 1980s, although with a different focus \cite{kleinert1982duality,kleinert1983double} (for a review see \cite{Kleinert,Beekman2017rev}). More recently various extensions of the original elastic dualities have been proposed. These include Cosserat elasticity with antisymmetric degrees of freedom \cite{Gromov:2019waa,Radzihovsky:2019jdo,Qi:2020jrf,Hirono:2021lmd}, elasticity with smectic anisotropy \cite{Radzihovsky_Smectic}, vortex lattices \cite{Nguyen:2020yve}, elasticity of quasicrystals \cite{Surowka:2021ved} and elasticity with underlying moir\'e lattices \cite{Gaa}. In addition, folding and tearing can have an interpretation as being fractonic \cite{Manoj:2020bcz}.

\subsection{Cauchy elasticity}
In its simplest incarnation the theory of elasticity consists of one displacement field $u_i$ that corresponds to the distortion of the underlying lattice. In general this field can have singularities that correspond to plastic deformations, generated by the underpinning topological defects. Our starting point is to construct an effective field theory that captures the dynamics of the displacement field $u_i$. In order to construct an action functional we note that the macroscopic field originates from the expectation value of the spontaneously broken translation symmetries represented by the position of the atoms in the lattice. Therefore the effective field theory shall not depend on the field itself but only its gradients. To the quadratic order in the fields we can write
\be
S[u^i] \equiv \int dt d^2x \mathcal{L}[u^i]= \int dt d^2x \Big[\dot{u}_i \dot{u}_i  - \frac{1}{2}C^{ijkl} u_{ij} u_{kl}\Big]\,,
\ee
where $C^{ijkl}$ is a tensor of elastic moduli and we introduce the symmetric strain tensor $u_{ij} = \partial_i u_j + \partial_j u_i$. We employ the Einstein summation convention. The resulting partition function is given by
\be
Z=\int Du^i e^{S[u^i]}\,.
\ee
This is the starting point of the fracton-elasticity duality. Next we change the elastic fields in favour of new collective bosonic fields by means of suitable Hubbard-Stratonovich transformations. Such transformations take the following form for given sets of fields $\phi $ and $\psi$
\be
\exp \left[\frac{1}{2} {\psi M \psi} \right]=\frac{1}{\mathcal{N}}   \int D \phi \exp \left[{\frac{-1}{2}  \phi M^{-1} \phi + \psi \phi} \right] \,.
\ee
In order to perform the Hubbard-Stratonovich transformation for elasticity it is natural to choose the canonically conjugate fields that correspond to momentum  $ P^i=T^{i0}$ and the stress tensor $T^{ij}$. The stress tensor is given by
\be
T^{ij} = -\frac{\delta \mathcal{L}}{\delta u_{kl} }= C^{ijkl} u_{kl}
\ee
One can express the original fields $u_{ij}$ in terms of the stress variables by inverting the four tensor $C^{ijkl}$. In the stress variables the action takes the following form
\be \label{eq:actionelast}
S[P^{i}, T^{ij}, u^i] = \int dt d^2x \Big[ P_i P^i + \mathcal{C}_{ijkl}T^{ij}T^{kl}+ 2u_i(\partial_\mu T^{i\mu})\Big]\,,
\ee
where we have introduced an inverse tensor of elastic coefficients $\mathcal{C}_{klmn}$ such that $C^{ijkl}\mathcal{C}_{klmn}= \text{Id}_{ijmn}=\delta_{im}\delta_{jn}$. The Hubbard-Stratonovich transformation doubles the degrees of freedom in the partition function. In the next step we want to remedy this by integrating out original fields $u_i$. 
\be \label{eq:dualnotresolved} 
Z = \int D P^{i}D T^{ij}D u^i e^{iS[P^{i}, T^{ij}, u^i]} = \int DP^{i}DT^{ij} e^{iS[P^{i}, T^{ij}]}\delta\left(\partial_\mu T^{i\mu}\right)\,,
\ee 
The Greek indices include both time and space coordinates. The delta function gives us a constraint that we would like to resolve by an appropriate choice of $T^{ij}$. This can be done by using gauge fields. However, $T^{ij}$ is a two-index object so we cannot resolve the delta function with the $U(1)$ gauge fields. We need fields $A_{ij}$ with two indices, dubbed tensor gauge fields. They are symmetric with respect to the permutations of indices. In terms of this gauge fields the stress tensor reads
 \be \label{eq:stressgauge}
T^{i\mu} = \epsilon^{\mu\nu\rho}\partial_\nu A^i_\rho\,.
\ee
Due to the antisymmetry of the Levi-Civita symbol the condition $\partial_\mu T^{i\mu}$ is always satisfied. In analogy with Maxwell electrodynamics we can define electric and magnetic fields
\begin{equation}
B^i =\epsilon^{kl}\partial_k A^i_l\,, \qquad E^i_j = \epsilon^i{}_k(-\partial_0A^k_j+\partial_j \partial _k A_0)\,.
\end{equation}
This allows one to write the dual theory \eqref{eq:dualnotresolved} in a gauge invariant manner
\be
\label{eq:actiondual}
S_{\text{dual}}= \int dt d^2x \frac{1}{2} \Bigg[ B_i B^i  +E_{ij}  \tilde{\mathcal{C}}^{ijkl}  E_{kl}\Bigg].\\ 
\ee
Tilde denotes index rotations, e.g. $\tilde{\mathcal{C}}_{ijkl}= \epsilon_{ii'} \epsilon_{jj'}\epsilon_{kk'}\epsilon_{ll'}\mathcal{C}^{i'j'k'l'}$. Fields $B_i$ and $E_{kl}$ are invariant under the following gauge transformations
\begin{equation} 
\delta A_{ij} = \partial_i \partial_j \alpha\,, \qquad \delta A_0 = \partial_0{\alpha}\,.
\end{equation}
We have shown that the theory of elasticity is in fact a tensor gauge theory. A natural question that emerges is what are the sources in this gauge theory, that we can introduce through a minimal coupling
\begin{equation}\label{eq:minimal}
\delta S =  \int dt d^2x \,\,\, \Big[\rho \delta A_0+ J^{ij}\delta A_{ij}\Big]\,
\end{equation}
 Since the gauge fields correspond to stresses in elasticity we know that they are generated by topological defects. As a result it is natural to expect that the topological defects are mapped to charges in the gauge theory. In order to see this mapping explicitely one can decompose the phonon into regular and singular parts $u^i = u^i_{\rm reg} + u^i_{\rm sing}$. Phonon displacement singularities couple to the conservation of the stress tensor
\begin{equation} \label{eq:sings}
\delta S = \int dt d^2x \,\,\, \Big[u^i_{\rm sing}\partial_\mu T^{i\mu}\Big].
\end{equation}
Comparing \eqref{eq:minimal} and \eqref{eq:sings} one can use Eq. \eqref{eq:stressgauge} and after integration by parts the mapping between defects and charges can be made explicit $\rho=\partial^i \rho_i$, $\rho^i = \epsilon^{i}{}_j\epsilon^{kl}\partial_k\partial_l u^{j}_{\rm sing}$ and $J^{ij} =\epsilon^{i}{}_{k} \epsilon^{\mu\nu j}\partial_\mu\partial_\nu u^k_{\rm sing}$.  The charge $\rho$ is mapped to the disclination density
\begin{equation}
\rho_{\rm disc} =  \frac{1}{2} \epsilon^k{}_{l}  \epsilon^{ij}\partial_i\partial_j \partial_k u^l_{\rm sing}\,.
\end{equation}

This concludes the basic features of elastic dualities. The main feature consists of the fact that elastic theories can be reformulated as gauge theories, whose charges are elastic defects. The duality introduced in this section constitutes the simplest example of a symmetric elasticity dual to a symmetric tensor gauge theory. Several more general theories of elasticity exist, constructed in order to describe systems  with additional degrees of freedom. We will briefly discuss some of such extensions, in which the dual fractonic theories have been constructed.

\subsection{Quasicrystal elasticity}
Symmetric elasticity introduced in the previous section is a macroscopic description of atoms localised periodically in space. This means that a discrete translation of the whole lattice is a symmetry of the system. The breaking of translation symmetry can be achieved in two different ways: either the localization of atoms becomes random or it preserves a quasi-periodic pattern. Quasicrystals are precisely crystals, in which the position of atoms is not arbitrary but the translation symmetry is broken \cite{levine1984quasicrystals,levine1985elasticity,socolar1986phonons}. In order to understand the elastic description of such crystals we consider a one-dimensional line of atoms parameterized by two sublattices $a$ and $b$. 
\be
\rho(x) = \sum_{n_a,n_b} [m \delta(x- n_a l_a)+m \delta(x-n_b l_b)],
\ee
where we have chosen the origin at $x=0$ and fixed $l_a$ and $l_b$ to be the averege separations of two lattices, whose ratio $l_a/l_b$ is an irrational number. $\rho(x)$ is a microscopic fluctuating field, whose average $\langle\rho(x)\rangle$ can be expressed as a discrete Fourier transform
\be
\langle\rho(x)\rangle = \sum_{p,q} \langle\rho_{pq}(x)\rangle\exp(i p k_a x+i q k_b x)\equiv \sum_{G}\langle\rho_{G}(x)\rangle\exp(i G_a x+i G_b x),
\ee
where the reciprocal lattice spanned by $G_a$ and $G_b$ is parameterized by $k_a=2\pi/l_a$ and $k_b=2\pi/l_b$ with $p,q\in\mathbb{Z}$. In analogy with the periodic case $\langle\rho_{G}(x)\rangle$ is a complex number with space-dependent amplitude and a phase
\be
\langle\rho_{G}(x)\rangle = \left|\langle\rho_{G}(x)\rangle\right|\exp[i G_a u_a(x)+i G_b u_b(x)+i\phi_o(x)]\equiv \left|\langle\rho_{G}(x)\rangle\right|\exp[i \phi_G(x)].
\ee
$u_a$ and $u_b$ describe displacements of the two sublattices. We can introduce new variables $u_a=u-w/2$ and $u_b=u+w/2$. $u=(u_a+u_b)/2$ corresponds to the phonon displacement and $w=u_b-u_a$ is known as the phason displacement that describes relative displacement of the two sublattices.

One can generalize the above argument to higher dimensions leading to $d$ component displacements vectors $u_i$ and $w_i$, where $d$ is the dimension of the lattice. In order to construct the elastic duality corresponding to quasi-crystals we focus on $d=2$. Our first step is to provide the action of the elastic fields in a quasi-crystal \cite{Ding1993}. The exact form is in general complicated but we can expand it in the fields $u_i$ and $w_i$. We note that the uniform displacements do not change the energy so the first terms in the expansion will be gradients of the fields. The phonon displacement leads to the symmetric strain tensor $u_{ij}=u_{ji}$, where $u_{ij}=\frac{1}{2}(\partial _i u_j+\partial_j u_i).$ The antisymmetric part corresponds to the rigid rotations of sub-lattices, which are not physically relevant. Contrary to the phonon field $u_{ij}$ the phason displacement tensor $w_{ij}=\partial _i w_j$ is not symmetric $w_{ij}\neq w_{ji}$. Physically the antisymmetric part is a consequence of the relative rotations of sub-lattices that cannot be neglected. The action can be written as a sum of the kinetic and potential energies $S[u_i,w_i]=S_{\text{kin}} [u_i,w_i]+S_{\text{pot}} [u_i,w_i]$, where the kinetic energy reads
\begin{equation}
S_{\text{kin}} [u_i,w_i] = \int dt d^2x \Big[\dot{u}_i \dot{u}_i  +\dot{w}_i \dot{w}_i\Big]\,,
\end{equation}
and the potential energy is given by
\begin{align}
\label{eq:freeen}
S_{\text{pot}} [u_i,w_i]&=- \frac{1}{2} \int dt d^2x \left[ \Big( C^{ijkl} u_{ij} u_{kl}\Big)\,\nonumber + \Big( K^{ijkl} w_{ij} w_{kl}\Big)+ \Big( R^{ijkl} w_{ij} u_{kl} + R'^{ijkl} w_{ij} u_{kl}\Big)\right]\\
&  \equiv -\frac{1}{2} \int  dt d^2x   \Big[ \left( u_{ij} \, w_{ij} \right) \begin{pmatrix} C_{ijkl} & R_{ijkl} \\R'_{ijkl}  & K_{ijkl}  \end{pmatrix} \begin{pmatrix}  u_{kl} \\ w_{kl} \end{pmatrix}  \Big]\, ,
\end{align}
where $R'_{klij}=R_{ijkl}$. This action can be written in the dual form, introducing the stress fields by varying the Lagrangian with respect to the stresses $\tilde{T}_{ij}=-\frac{\partial \mathcal{L}}{\partial u_{ij}}$, $H_{ij}=-\frac{\partial \mathcal{L}}{\partial w_{ij}}$,
and subsequently rewriting them using gauge potentials $\tilde{T}^{i\mu} = \epsilon^{\mu\nu\rho}\partial_\nu A^i_\rho\,$, $H^{i\mu} = \epsilon^{\mu\nu\rho}\partial_\nu \mathcal{A}^i_\rho\,$, and the corresponding electric and magnetic fields
\begin{align}
\label{eq:actiondual}
S_{\text{dual}}&= \int dt d^2x \frac{1}{2} \Bigg[ B_i B^i + \mathcal{B}_i \mathcal{B}^i\,\nonumber  +\left( E_{ij} \, \mathcal{E}_{ij} \right) \begin{pmatrix} \tilde{\mathcal{C}}_{ijkl} & \tilde{\mathcal{R}}_{ijkl} \\ \tilde{\mathcal{R}}'_{ijkl}  & \tilde{\mathcal{K}}_{ijkl}  \end{pmatrix} \begin{pmatrix}  E_{kl} \\ \mathcal{E}_{kl} \end{pmatrix} \Bigg].
\end{align}
The fields are defined in the following way
\begin{equation}
B^i =\epsilon^{kl}\partial_k A^i_l\,, \qquad E^i_j = \epsilon^i{}_k(-\partial_0A^k_j+\partial_j \partial _k A_0)\,.
\end{equation}
\begin{equation}
\mathcal{B}^i =\epsilon^{kl}\partial_k \mathcal{A}^i_l\,, \qquad \mathcal{E}^i_j = \epsilon^i{}_k(-\partial_0 \mathcal{A}^k_j+\partial_j\mathcal{A}^k_0)\,.
\end{equation}
These fields are invariant under the following gauge transformations
\begin{equation} 
\delta A_{ij} = \partial_i \partial_j \alpha\,, \qquad \delta A_0 = \partial_0{\alpha}\,,
\end{equation}
\begin{equation}
\delta \mathcal{A}_{ij}= \partial_j \beta_i \, , \qquad \delta\mathcal{A}_{i0} = \partial_0{\beta}_i\,.
\end{equation}
We can now source the gauge fields by appropriate sources
\begin{equation}
\mathcal{L}_{\text{sources}}=  A_0 \rho +A_{ij}J_{ij}+ \mathcal{A}_{i0}  \varrho_i+\mathcal{A}_{ij}\mathcal{J}_{ij} \,.
\end{equation}
In addition to the dislocation defects that we already saw in the symmetric elasticity a new type of defects can appear in quasi-crystals - matching or stacking faults. These are singularities in the phason field leading to a non-zero charge density
\begin{equation}
\varrho^i = \epsilon^{i}{}_j\epsilon^{kl}\partial_k\partial_l w^{j}_{\rm sing}
\end{equation}
and the current
\begin{equation}
\mathcal{J}^{ij} =\epsilon^{i}{}_{n} \epsilon^{\mu\nu j}\partial_\mu\partial_\nu w^n_{\rm sing}.\,
\end{equation}
We can now identify the duality mapping between charges and defects for phasons. Vector charges in the dual theory map to the rotated matching faults $ \epsilon^{i}{}_j \varrho^j$.

\subsection{Cosserat elasticity}
Symmetric elasticity assumes that the solid constituents have no internal structure. However, in complex materials, composed from elongated bodies this assumption is not valid and one has to consider rotational degrees of freedom, again generalizing the macroscopic description of a solid \cite{Nowacki,Eringen,Sadd}. We want to write the effective action describing such a solid. In the first step we need to identify the degrees of freedom. In two dimensions the displacement vector $u_i$ is supplemented with an orientation angle $\theta$. In the second step we require that the effective action is invariant under translations and rotations. Translations require that under the transformation $u_i\rightarrow u_i+b_i$, where $b_i$ is a constant vector the action remains invariant. Rotations by a constant angle $\theta_0$ are implemented by two simultaneous transformations $\theta \rightarrow \theta +\theta_0$, $u_i\rightarrow u_i+ \epsilon_{ij} x_j \theta_0$. We note that gradients of the displacement field are invariant under translations but not under rotations. It is, however, possible to construct a combination
\be
\gamma_{ij} = \partial_i u_j - \epsilon_{ij}\theta\,,
\ee
that is invariant both under translations and rotations. Using this invariant one can write down the quadratic form of the effective action
\be
S[u^i,\theta] = \int dt d^2x \Big[ \dot{\theta}\dot{\theta} + \dot{u}^i \dot{u}^i - C^{ijkl}\gamma_{ij}\gamma_{kl} + \zeta\tau_i \tau^i \Big]\,,
\ee
where $C^{ijkl}$ and $\zeta$ denote elastic coefficients in the theory and $\tau_i =\partial _i\theta$. By writing the action in terms of $u_i$ and $\theta$ we can immediately conclude that we have massive modes in the theory. As a result orientational modes are not massless Goldstone modes. This is a general feature of the spontaneous breaking of spacetime symmetries that leads to a smaller amount of massless modes than the symmetries. This also leads to a complication in order to directly implement the duality transformation. Performing the Hubbard-Stratonovich transformation
\be
S = \int dt d^2x \Big[ P_iP^i + (L_0)^2 + \zeta^{-1}L_iL^i + \mathcal{C}_{ijkl}T^{ij}T^{kl} + u_i\Big(\partial_\mu T^{i\mu}\Big) + \theta\Big(\partial_\mu L^\mu - \epsilon^{ij}T_{ij}\Big) \Big],
\ee
and integrating out the smooth part of $\theta$ and $u_i$ leads to the following constraints
\be\label{eq:consCosserat}
\partial_\mu T^{i\mu}=0\,, \qquad \partial_\mu L^\mu - \epsilon^{ij}T_{ij}=0\,.
\ee
The second constraint does not have a form of a conservation law. Therefore we need to first resolve the first constraint as in \eqref{eq:stressgauge}, express $T_{ij}$ as a gradient and only then resolve it the second constraint using an ordinary $U(1)$ gauge field
\be
L^0 + \epsilon^{ij}A_{ij} = \epsilon^{ij}\partial_i a_j=b\,, \qquad L^i + \epsilon^{ij}A_{j0} = \epsilon^{ij} (\partial_ia_0-\partial_0 a_i) = \epsilon^{ij}e_j\,,
\ee
The action for the dual gauge fields takes form
\be\label{eq:dualCosserat}
S = \int dt d^2x \Big[ \tilde{ \mathcal{C}}^{ijkl}E_{ij}E_{kl} + B_i B^i + \zeta^{-1}(b+\epsilon^{ij}A_{ij})^2 + ( e^i -A^i_0)( e_i - A_{i0})\Big]\,.
\ee
It is invariant under the following set of transformations
\bea \label{eq:gauget1}
&&\delta a_\mu = \partial_\mu \lambda 
\\ \label{eq:gauget2}
&&\delta A_{i0} = \partial_0{\alpha}_i\,, \qquad \delta A_{ij}= \partial_j \alpha_i\,, \qquad \delta a_i =-\alpha_i\,, \qquad \delta a_0 = 0.
\eea
In the final step we show the charges of the dual theory that again correspond to defects in Cosserat elasticity
\be
\mathcal{L}_{\text{sources}}= \rho^i_{\text{rot}} A_{i0} +J_{\text{rot}}^{ij} A_{ij}+  a_0 \rho_\theta +a_{i}j_{i}= \Big[ (\rho^i +2 \epsilon^{ij}\partial_j \theta_{\rm sing})A_{i0} + (J^{ij} + 2\dot{\theta}_{\rm sing}\epsilon^{ij})A_{ij} +  a_0 \rho_\theta +a_{i}j_{i} \Big]\,,
\ee
where $ \rho_\theta =\epsilon^{ik}\partial_k\partial_i \theta_{\rm sing}$, $j^i = \epsilon^{ik}\left(\partial_k \partial_0 - \partial_0 \partial_k\right)\theta_{\rm sing}\,.$ We note that the dislocation density have contributions from singularities of both the displacement and orientation fields. As such, in order to correctly account for fractonic behavior of defects, in Cosserat theory both contributions have to be taken into account.


\section{Effective theories and fracton hydrodynamics}  \label{sec:hydro}

We have seen how tensor gauge theories emerge in the context of elasticity and couple to fracton excitations corresponding to lattice defects. However, the dynamics of the fracton themselves is not captured by the tensor fields. Developing effective field theories of fractons is interesting both for conceptual reasons and for their possible application to the description of the collective behavior of fractons. An instance relevant for physical systems is the quantum melting of solids, where the transition is produced by the formation of a fracton condensate \cite{Pretko:2018tit,Kumar_2019,Zhai:2019duk,Nguyen:2020yve}.

An early proposal for a fracton effective theory was given in the context of the X-cube model of fracton topological order by Slagle and Kim \cite{Slagle:2017wrc}. This was generalized by Pretko taking a global vector symmetry as guiding principle \cite{Pretko:2018jbi}, and further generalizations by Seiberg followed up a bit later \cite{Seiberg:2019vrp}. Fracton condensation is only mentioned in passing in Pretko's work. More detailed studies of the properties of phases with a fracton condensate (``fracton superfluids'') have been made by Chen, Ye and Yuan \cite{Yuan:2019geh,Chen:2020jew}, that also generalized Pretko's action by including anisotropic terms.

The main idea is simple. A collective system of spinless bosons may be described by a complex scalar $\psi$ with a global $U(1)$ symmetry that corresponds to the conserved particle number. The $U(1)$ transformation of the field is the usual one
\begin{equation}
\psi\to e^{i\alpha} \psi.
\end{equation}
If instead of ordinary bosons one has a system of fractons with conserved dipole charge, then the $U(1)$ symmetry is enhanced to a global vector symmetry
\begin{equation}
\psi\to e^{i\boldsymbol\beta\cdot \boldsymbol x} \psi,
\end{equation}
where $\boldsymbol \beta$ is a constant vector along the spatial directions $\boldsymbol x$. Imposing this symmetry constrains the action of the scalar field, in particular it forbids an ordinary kinetic term with spatial derivatives. A possible Lagrangian density to quartic order in the field is \cite{Pretko:2018jbi,Seiberg:2019vrp}
\begin{equation}\label{eq:fracEFT}
\begin{split}
{\cal L}=&|\partial_0\psi|^2- m^2|\psi|^2-\frac{\lambda}{4}(|\psi|^2)^2\\
&-c_1\partial_i |\psi|^2 \partial_i |\psi|^2-c_2|\psi\partial_i \partial_j \psi-\partial_i \psi\partial_j \psi|^2-c_3 \left[(\psi^*)^2\left( \psi \partial^2 \psi-\partial_i \psi \partial_i \psi\right)+h.c.\right].
\end{split}
\end{equation}
The gauging of this theory has been considered by Banerjee \cite{Banerjee:2021lqe,Banerjee:2021qcj}. Remarkably, imposing a vector global symmetry in this way introduces an emergent {\em subsystem symmetry}  \cite{Pretko:2018jbi} around the trivial vacuum $\psi=0$. Indeed, the quadratic action for a perturbation around the trivial vacuum is
\begin{equation}
{\cal L}\simeq |\partial_0\delta \psi|^2- m^2|\delta\psi|^2.
\end{equation}
The mass term is removed by a field redefinition $\delta \psi=e^{i m t}\varphi$,
\begin{equation}
{\cal L}\simeq |\partial_0\varphi|^2-i m \left(\varphi^*\partial_0 \varphi-\partial_0\varphi^*\varphi \right).
\end{equation}
In this form, the quadratic Lagrangian changes by a total derivative under the subsystem transformations
\begin{equation}
\delta \varphi=\alpha(\boldsymbol x)\ \Rightarrow \ \delta{\cal L}=-im\partial_0 \left(\alpha^*\varphi- \varphi^*\alpha  \right).
\end{equation}
The emergent subsystem symmetry implies that excitations around the trivial vacuum are immobile, they would also be gapless if the mass vanishes $m=0$. Note that quantum corrections may modify the gap, but as long as the vector global symmetry is not anomalous, they would not introduce a kinetic term, so that excitations remain immobile. It should be noted that emergent subsystem symmetries and fractonic dispersion relations are not limited to theories with a global vector symmetry (or other polynomial shift symmetries), they may also appear in other situations, such as the model with spontaneously broken translation invariance studied by Argurio, Hoyos, Musso and Naegels \cite{Argurio:2021opr}. Exact subsystem symmetries can be realized in a variety of continuum theories \cite{slagle2019foliated,slaglefoliated2,hsin2021comments,aasen2020topological,Seiberg:2020bhn,Seiberg:2020wsg,Seiberg:2020cxy,Gorantla:2020xap,Karch:2020yuy,shirley2020twisted,Ma:2020svo} (see \cite{Geng:2021cmq} for a nice summary). 

By tuning the potential to $m^2<0$, it is possible to change the ground state to a non-trivial vacuum $\psi=\psi_0$. This corresponds to a phase were fractons have condensed, or in other words a fracton superfluid. There is a gapless degree of freedom corresponding to the phase of the condensate, which can be identified as a Nambu-Goldstone mode
\begin{equation}
\psi=\psi_0 e^{i\phi}.
\end{equation}
In contrast to the fluctuations around the trivial vacuum, the action for fluctuations of the Nambu-Goldstone mode includes terms with spatial derivatives in the quadratic action 
\begin{equation}\label{eq:fractonSF}
{\cal L}\simeq |\psi_0|^2(\partial_0 \phi)^2-c_2(|\psi_0|^2)^2(\partial_i \partial_j \phi)^2-c_3(|\psi_0|^2)^2(\partial^2\phi)^2.
\end{equation}
In this case the global vector symmetry is realized as a polynomial shift symmetry
\begin{equation}
\delta \phi= \boldsymbol\beta \cdot \boldsymbol x,
\end{equation}
but there is no emergent subsystem symmetry, so there are no immobile fluctuations in the condensed fractonic phase. The same is true in more involved cases \cite{Yuan:2019geh,Chen:2020jew}. One should note that \eqref{eq:fractonSF} just captures the semiclassical description of the fracton superfluid, the robustness of the superfluid state to fluctuations has been studied by Stahl, Lake and Nandkishore \cite{Stahl:2021sgi}.

Instead of a fracton condensate one might consider a state with thermally excited fractons. Let us assume that the system reaches thermal equilibrium and can be well approximated by a homogeneous state in the effective theory description. After the state is perturbed, the late time relaxation to the equilibrated state should be captured by hydrodynamics, in the general sense of a theory describing transport of conserved charges at long wavelengths. A first proposal of a transport theory in the hydrodynamic regime with a conserved charge density was made by Gromov, Lucas and Nandkishore \cite{Gromov:2020yoc}.

In fractonic systems we expect to either have dipole or higher moment conserved charges and/or subsystem symmetries. Let us assume that moments up to order $n$ are conserved. Denoting $\rho$ as the fracton number density, the associated charges in a $d+1$-dimensional theory are
\begin{equation}
Q^{(k)}[\boldsymbol a]=\int d^d x\, a_{i_1\cdots i_k}x^{i_1}\cdots x^{i_k} \, \rho.
\end{equation}
For instance, an often discussed case that we will keep as an example in the discussion is when the dipole moment is conserved in any direction and the second ``trace'' moment $\sim {\boldsymbol x^2} $ is conserved as well. 

Conservation of the fracton number, dipole and higher moment charges can be achieved if the density satisfies a conservation law of the following form
\begin{equation}
\partial_0 \rho +\partial_{i_1}\cdots \partial_{i_{n+1}} J^{i_1\cdots i_{n+1}}=0.
\end{equation}
Here $J^{i_1\cdots i_{n+1}}$ is a completely symmetric tensor current. In some cases the conservation equation may involve a lower number of spatial derivatives as long as the tensor current satisfies some algebraic conditions. For instance, in the example we are considering, the dipole and second moment current are conserved for
\begin{equation}
\partial_0 \rho+\partial_i \partial_j J^{ij}=0,\ \ J^i_{\ i}=0.
\end{equation} 
In this case there are two rather than three spatial derivatives in the conservation equation, but the second moment is conserved thanks to the tracelessness condition that the current satisfies. In general, we expect that the number of spatial derivatives in the conservation equation will be determined by the highest irreducible representation (the highest multipole) under spatial rotations among the conserved charges.

In the most general situation, spatial momentum may not be conserved and the hydrodynamic equations reduce to the continuity equation for the fracton density and energy conservation. In the hydrodynamic effective description the only dynamical degrees of freedom are the conserved charges. The currents are determined by the densities through the constitutive relations, if the $n$-multipole is the highest conserved, the current must take the form
\begin{equation}
J^{i_1\cdots i_{n+1}}=\frac{D}{N}\partial^{i_1}\cdots \partial^{i_{n+1}}\rho -(\textrm{all traces}).
\end{equation}
Where by $N$ we have denoted the number of all possible terms in the constitutive relation. This leads to an equation for the density
\begin{equation}
\partial_0 \rho+D \left(\partial^2\right)^{n+1} \rho=0.
\end{equation}
Therefore, the system is diffusive but compared to ordinary diffusion where the spatial width of a distribution increases as a square root of time $\Delta x\sim t^{1/2}$, the conservation of higher multipole charges slows down the process to $\Delta x\sim t^{1/(n+1)}$, i.e. it is {\em subdiffusive} \cite{Gromov:2020yoc,2020PhRvB.101u4205M,Feldmeier:2020xxb,Moudgalya:2020mpa,2020arXiv200408695Z,2020PhRvX..10a1042G,Ganesan:2020wvm,2021PhRvE.103b2142I}.

If rotational invariance is broken, then the higher moment symmetries can be further enhanced to subsystem symmetries where the fracton density multiplied by an arbitrary function of some of the coordinates corresponds to a conserved charge. A simple example is a two-tensor current that along one direction has only off-diagonal components, say $J^{11}=0$. Then, the following charges are conserved
\begin{equation}
Q_1[f]=\int d^dx f(x^1) \rho.
\end{equation}
This implies a degeneracy in the continuity equation that leads to degeneracy in the dispersion relations and fractonic behaviour. In our example the current must take the form $J^{1i}=J^{i1}=-D_1\partial_1 \partial_i\rho$, $J^{ij}=-D_2\partial_i\partial_j\rho$, $i,j\neq 1$. Then, solving the equation using a Fourier transform, one finds the subdiffusive mode
\begin{equation}
\omega=-i{\boldsymbol k}_\perp^2 (2 D_1 k_1^2+D_2 {\boldsymbol k}_\perp^2),\ \ {\boldsymbol k}_\perp^2=\sum_{i\neq 1} k_i^2. 
\end{equation}
The dispersion relation vanishes along the $k_1$ direction at the origin of the transverse space ${\boldsymbol k}_\perp^2=0$.

The effective theory \eqref{eq:fracEFT} is invariant under translations, so that, in addition to the dipole charge, spatial momentum is conserved. Ideal fracton hydrodynamics with dipole and spatial momentum conservation was introduced  by Grosvenor, Hoyos, Pe\~na-Benitez and Sur\'owka \cite{Grosvenor:2021rrt} and, in the work of Glorioso, Guo, Rodriguez-Nieva, Lucas \cite{Glorioso:2021bif}, dissipative terms were also taken into account, introducing the characteristic subdiffusive behavior\footnote{See also the more recent work \cite{Osborne:2021mej}.}. At the ideal level there are gapless propagating modes with quadratic dispersion relation, in contrast to the linear dispersion of ordinary phonons, and quite similar to the Nambu-Goldstone modes of  \eqref{eq:fractonSF}. A more involved case with a chiral fluid and subsystem symmetry was also studied, but its behaviour does not follow this simple picture. Nevertheless, in both cases it was shown that hydrodynamic equations are essentially the same as those of an ordinary fluid, with the higher moment or subsystem charge conservation following from the constitutive relations of the particle number current. This is reminiscent of other coordinate-dependent symmetries like scale or conformal invariance.

\section{Conclusions}

Motivated by the exotic properties of fractons, a rich set of theories has been uncovered, many of them realized in lattices and other condensed matter systems. Multipole and other symmetries with transformations depending on spatial coordinates may serve as a guiding principle to construct and classify these novel theories in the continuum limit, as we have tried to emphasize in this short overview. 

Many questions remain pertaining the significance of gauge symmetries of tensor fields. Eminently, it is unclear whether they should be treated on equal footing with ordinary or higher form gauge symmetries, or rather physical states do not need to be gauge invariant under tensor symmetries. The difference between form and tensor gauge symmetries is highlighted when attempting to couple each type of theory to a curved background geometry. While form gauge transformations remain metric-independent and are trivially maintained (barring quantum anomalies), this is not obviously so for tensor symmetries where in principle covariant derivatives would appear in the gauge transformation \footnote{Recently, this was addressed in \cite{Bidussi:2021nmp,Jain:2021ibh} using Aristotelian geometry.}. 
The Heisenberg group construction that allowed a consistent coupling of dipole symmetric theories to a background geometry \cite{Pena-Benitez:2021ipo} could be used as guidance for more general coordinate-dependent symmetries.

Other aspect that has been not explored in detail beyond lattice models is the dynamics of fracton excitations, either individual or collective. In the paradigmatic example of elasticity, fractons are defects in the crystal lattice, and from this point of view they correspond to singular configurations of displacement fields, and perhaps not amenable to treatment in a continuum theory. An analog of this situation is that of vortices in a superfluid, which are singular configurations of the phase of the condensate. Nevertheless, a collective description of superfluid vortices can be captured by a continuum effective theory (see e.g.~\cite{Watanabe:2013iia,Moroz:2018noc}) and possibly similar derivations could be obtained for theories with fractonic excitations beyond the simple examples shown here.  

Summarizing, this is a rapidly evolving field with many open avenues where we expect to witness significant progress in the near future.

\section*{Acknowledgements}

K.T.G. has received funding from the European Union's Horizon 2020 research and innovation programme under the Marie Sk\l odowska-Curie grant agreement No 101024967. C.~H. has been partially supported by the Spanish Ministerio de Ciencia, Innovaci\'on y Universidades through the grant PGC2018-096894-B-100. P.S. was supported by the Deutsche Forschungsgemeinschaft through the Leibniz Program, and the Narodowe Centrum Nauki (NCN) Sonata Bis grant 2019/34/E/ST3/00405. F.P-B. has received funding from the Norwegian Financial Mechanism 2014-2021 via the Narodowe Centrum Nauki (NCN) POLS grant 2020/37/K/ST3/03390. P.S and F.P-B. were supported from  the cluster of excellence ct.qmat (EXC 2147, project-id 39085490).


\bibliographystyle{utphys2}
\bibliography{FractonReview}

\end{document}